\input{epsf} 
\newcommand{\et}{\hspace{-0.08in}{\bf .}\hspace{0.1in}}
\newcommand{\BOX}{\hbox {$\sqcap$ \kern -1em $\sqcup$}}

\newcommand{\R}{{\Bbb R}}
\newcommand{\g}{{\frak g}}
\newcommand{\so}{{\frak so}}

\renewcommand{\to}{\rightarrow}

\newcommand{\maps}{\colon}
\newcommand{\eps}{\epsilon}
\newcommand{\iso}{\cong}

\newcommand{\tr}{{\rm tr}}

\newcommand{\SU}{{\rm SU}}

\newcommand{\SO}{{\rm SO}}

\newcommand{\Ad}{{\rm Ad}}

\newcommand{\T}{{\cal T}}
\newcommand{\we}{\wedge}
\newtheorem{thm}{Theorem}    

        \newcommand{\be}{\begin{equation}}
        \newcommand{\ee}{\end{equation}}
        \newcommand{\ba}{\begin{eqnarray}}
        \newcommand{\ea}{\end{eqnarray}}
        \newcommand{\ban}{\begin{eqnarray*}}
        \newcommand{\ean}{\end{eqnarray*}}
        \newcommand{\barr}{\begin{array}} 
        \newcommand{\earr}{\end{array}}

\documentstyle[12pt,amsfonts,diagram]{article}
\begin{document}

      \begin{center}
      {\bf Degenerate Solutions of General Relativity \\
       from Topological Field Theory \\}
      \vspace{0.5cm}
       \vspace{0.5 cm}
      {\em John C.\ Baez\\}
      \vspace{0.3cm}
      {\small Department of Mathematics,  University of California\\ 
      Riverside, California 92521 \\
      USA\\ }
      \vspace{0.3cm}
      {\small email: baez@math.ucr.edu\\}
      \vspace{0.3cm}
      {\small February 24, 1997 \\ }
      \end{center}

\begin{abstract}
\noindent 
Working in the Palatini formalism, we describe a procedure for constructing
degenerate solutions of general relativity on 4-manifold $M$ from
certain solutions of 2-dimensional $BF$ theory on  any framed surface
$\Sigma$ embedded in $M$.  In these solutions the cotetrad field $e$
(and thus the metric) vanishes outside a neighborhood of $\Sigma$,
while inside this neighborhood the connection $A$ and the field $E = e
\we e$ satisfy the equations of 4-dimensional $BF$ theory.  Moreover,
there is a correspondence between these solutions and certain solutions
of $2$-dimensional $BF$ theory on $\Sigma$.  Our 
construction works in any signature and with any value of the
cosmological constant.   If $M = \R \times S$ for some 3-manifold $S$,
at fixed time our solutions typically describe `flux tubes of area':
the 3-metric vanishes outside a collection of thickened links embedded
in $S$, while inside these thickened links it is nondegenerate only
in the two transverse directions.  We comment on the quantization of the
theory of solutions of this form and its relation to the loop
representation of quantum gravity.

\end{abstract}

\section{Introduction}

Since the introduction by Rovelli and Smolin of spin networks into
what used to be called the `loop representation' of quantum
gravity, progress on understanding the kinematical aspects of the
theory has been swift \cite{A,B4,RS}.  There is a basis of
kinematical states given by spin networks, i.e., graphs embedded in
space with edges labelled by irreducible representations of
$\SU(2)$, or spins, and with vertices labelled by intertwining
operators.  Geometrical observables have been quantized and their
matrix elements computed in the spin network basis, giving a
concrete --- though of course still tentative --- picture of
`quantum 3-geometries'.  In this picture, the edges of spin networks
play the role of quantized flux tubes of area: surfaces
acquire area through their intersection with these edges, each edge
labelled with spin $j$ contributing an area equal to $\sqrt{j(j+1)}$
times the Planck length squared for each transverse intersection
point \cite{AL,RS2}.

Unsurprisingly, the dynamical aspects remain more obscure.  Thiemann
has proposed an explicit formula for the Hamiltonian constraint
\cite{T3}, but this remains controversial, in part because of the
difficulty in extracting physical predictions that would test this
proposal, or any competing proposals.  A crucial problem is the lack
of a clear picture of the {\it 4-dimensional}, or {\it spacetime},
aspects of the theory.  For example, diffeomorphism equivalence
classes of states annihilated by the Hamiltonian constraint should
presumably represent physical states of quantum gravity.  Many
explicit solutions are known, perhaps the simplest being the `loop
states', corresponding to spin networks with no vertices, or in other
words, links with components labelled by spins.  However, the
interpretation of these states as `quantum 4-geomeunclear.

In search of some insight into the 4-dimensional interpretation of
these loop states, we turn here to classical general relativity.  We
construct degenerate solutions of the equations of general relativity
which at a typical fixed time describe `flux tubes of area'
reminiscent of the loop states of quantum gravity.  More precisely,
the 3-metric vanishes outside a collection of embedded solid tori,
while inside any of these solid tori the metric is degenerate in the
longitudinal direction, but nondegenerate in the two transverse
directions.

The 4-dimensional picture is as follows: given any surface $\Sigma$
embedded in spacetime, we obtain solutions for which the metric
vanishes outside a tubular neighborhood of $\Sigma$.  Inside this
neighborhood, which we may think of as $\Sigma \times D^2$, the
4-metric is degenerate in the two directions tangent to $\Sigma$ but
nondegenerate in the two transverse directions.   In the 4-geometry 
defined by one of these solutions, the area of a typical surface
$\Sigma'$ intersecting $\Sigma$ transversally in isolated points
is determined by a sum of contributions from the point in the intersection
$\Sigma \cap \Sigma'$.  

The solutions we consider are inspired by the work of Reisenberger
\cite{R1}, who studied a solution for which the metric vanishes
outside a neighborhood of a 2-sphere.  Since various formulations of
Einstein's equation become inequivalent when extended to the case of
degenerate metrics, it is important to say which formulation is being
used in any work of this kind.  Reisenberger worked in the Plebanski
(or self-dual) formalism, and concentrated on the initial-value
problem for general relativity extended to possibly degenerate complex
metrics, comparing Ashtekar's version of the constraints and a
modified version which differs only when the densitized triad field
has rank less than $2$.  Only the latter version admits his `2-sphere
solution'.  We work 4-dimensionally for the most part, using the
classical equations of motion coming from the Palatini formalism, in
which the basic fields are a cotetrad field $e$ and a
metric-preserving connection $A$ on the `internal space' bundle.  Our
work treats arbitrary signatures and arbitrary values of the
cosmological constant.

An interesting aspect of our solutions is that they arise from
solutions of 2-dimensional $BF$ theory, a topological field theory, on
the surface $\Sigma$.  This takes advantage of the relation between
general relativity and $BF$ theory in 4 dimensions, together with the
fact that $BF$ theory behaves in a simple manner under dimensional
reduction.  

It is also interesting that our construction requires us to choose a
thickening of $\Sigma$, that is, to extend the embedding of $\Sigma$
in spacetime to an embedding of $\Sigma \times D^2$.  The topological
data needed to specify a thickening of $\Sigma$ up to diffeomorphism
is known as a `framing' of $\Sigma$.  This is precisely the
4-dimensional analog of a framing of a knot or link in a 3-manifold.
The possible need for framings in the loop representation of canonical
quantum gravity has been widely discussed \cite{B5,BGGP,MS}, but here 
we see it arising quite naturally in the classical context, where 
the spacetime aspects of framing-dependence are easier to 
understand.  

In the Conclusions we speculate on some of the possible
implications of our ideas for the quantum theory.  We discuss a
`minisuperspace' model in which one quantizes only the
degenerate solutions of general relativity associated to a fixed
surface $\Sigma$ in spacetime.   We treat this model in detail
in another paper \cite{B6}; here we only sketch the main ideas.
For simplicity we restrict our attention to the Lorentzian signature
and certain nonzero values of the cosmological constant $\Lambda$.  
Our main result is that this minisuperspace model is equivalent to a 
$G/G$ gauged Wess-Zumino-Witten model on the surface $\Sigma$, with 
gauge group $\SO(3)$.  It follows that if $\Sigma$ intersects space at 
a given time in some link $L$, states at this time correspond to labellings
of the components of $L$ with irreducible representations of quantum
$\SO(3)$, or in other words, integral spins 
\[    j = 0, 1, 2, \dots, k/2 \]
where the `level' $k$ is an even integer depending on $\Lambda$.  In
other words, our states are described by a special class of the 
`$q$-deformed spin networks' intensively studied by knot theorists
\cite{KL}.  Major and Smolin \cite{MS} have suggested using
$q$-deformed spin networks to describe states of 
quantum gravity with nonzero cosmological constant; here we
see quite concretely how $q$-deformed spin networks arise in a
toy model.

\section{General relativity and $BF$ theory}

Understanding the degenerate solutions of general relativity obtained
from topological field theory is perhaps simplest in the Palatini
formalism.  We begin by recalling some notation introduced in our
previous papers \cite{B2,B4}. 

In the Palatini formalism our spacetime $M$ can be any oriented
smooth 4-manifold equipped with an oriented vector bundle $\T$ that
is isomorphic to $TM$ and equipped with a (nondegenerate) metric $\eta$. 
In the rest of this section $\eta$ can have any fixed signature.  The
fiber of $\T$ is called the `internal space', and $\eta$ is the
`internal metric'.  The basic fields are a $\T$-valued 1-form
$e$ on $M$, often called the `soldering form' or `cotetrad field',
and a metric-preserving connection $A$ on $\T$.   
Pulling back the internal metric along $e \maps \T \to TM$ we obtain
a --- possibly degenerate --- metric $g$ on $M$, given explicitly by
\[   g(v,w) = \eta(e(v), e(w)) .\]
The metric $g$ is nondegenerate precisely when $e$ is an isomorphism.

To write the Palatini action in index-free notation,  it is handy to
work with differential forms on $M$ taking values in the exterior
algebra bundle $\Lambda \T$.  In these terms, the Palatini action
with cosmological constant $\Lambda$ is given by 
\be
S_{Pal}(A,e) = \int_M \tr(e \we e \we F + {\Lambda\over 12} e \we e \we
e \we e) 
\label{Palaction} \ee 
Here we use the internal metric to regard the curvature $F$ of
$A$ as a $\Lambda^2 \T$-valued 2-form on $M$, and use the symbol
`$\we$' to denote the wedge product of differential forms tensored
with the wedge product in $\Lambda \T$.   Also, the orientation and
internal metric on $\T$ give an `internal volume form', i.e., a
section of $\Lambda^4\T$, and thus a map from $\Lambda^4\T$-valued
forms to ordinary differential forms, which we denote above by `$\tr$'.  

If we set the variation of this action equal to zero,
ignoring boundary terms, we obtain the field equations
\be    d_A(e \we e) = 0,\qquad 
 e \we (F + {\Lambda \over 6} e \we e) = 0 
\label{einstein} \ee
where $d_A$ denotes the covariant exterior derivative.   If $e$ is an
isomorphism, the first equation implies that $d_A e = 0$, which means
that the connection on $TM$ corresponding to $A$ via $e$ is
torsion-free, and thus equal to the Levi-Civita connection of $g$. 
The latter equation is then equivalent to Einstein's equation for $g$. 
Therefore these equations describe an extension of general relativity to
the case of degenerate metrics.  

Now compare $BF$ theory.   This makes sense in any dimension, but
has special features in dimension 4.  First let $M$ be any
oriented smooth $n$-manifold and let $P$ be a $G$-bundle over $M$,
where $G$ is a connected semisimple Lie group.   Then the basic fields of the 
theory are a connection $A$ on $P$ and an $\Ad P$-valued 
$(n-2)$-form, often called $B$, which we shall call $E$
because of its relation to the gravitational electric field.
The action of the theory is
\[       \int_M \tr(E \we F)  ,\] 
where $F$ is the curvature of $A$ and
`$\tr$' is defined using the Killing form on $\g$.  
The corresponding field equations are
\[      d_A E = 0, \qquad F = 0  .\]
In dimension 4 we can also add a cosmological constant term to the
action, obtaining 
\be   S_{BF}(A,E)  = \int_M \tr (E \we F + {\Lambda \over 12} E \we E) .
\label{BFaction} \ee
which gives the field equations
\be       d_A E = 0 , \qquad  F + {\Lambda\over 6} E = 0. \label{BF} \ee
The case $\Lambda \ne 0$ is very different from the case
$\Lambda = 0$.  When $\Lambda \ne 0$, the second equation follows from
the first one and the Bianchi identity, so $A$ is arbitrary and it
determines $E$.  When $\Lambda = 0$, $A$ is flat and $E$ is
any solution of $d_A E = 0$.  

We can relate $BF$ theory in 4 dimensions to general relativity in the
Palatini formalism if we take $G = \SO(4), \SO(3,1),$ or $\SO(2,2)$,
depending on the signature of the internal metric, and we let $P$ be the
special orthogonal frame bundle of $\T$.  This lets us think of $E$ as
a $\Lambda^2 \T$-valued 2-form, just as $e \we e$ is  in the Palatini
formalism.  Comparing the $BF$ theory equations with those of 
general relativity, it is clear that $e \we e$ and $E$ play 
analogous roles.   The easiest way to exploit this is to note that if 
\[  E = e \we e, \]
then the equations of $BF$ theory, (\ref{BF}), imply those of
general relativity, (\ref{einstein}).  An analogous observation
has been widely remarked on in the context of the Ashtekar formalism
(see \cite{B2} and the many references therein).  Our new observation
here is that for (\ref{einstein}) to hold, it suffices for {\rm
(\ref{BF})} to hold where $e$ is nonzero.  In what follows we will be
interested in degenerate solutions of general relativity where $e$
vanishes everywhere except in a neighborhood of a 2-dimensional
surface $\Sigma$, and the equations of 4-dimensional $BF$ theory hold
in this neighborhood.  We construct these solutions from certain
solutions of 2-dimensional $BF$ theory living on $\Sigma$.

\section{Local analysis}

We begin by working locally in coordinates
$(t,x,y,z)$ and fixing a local trivialization of $\T$.  
We consider the surface 
\[         \Sigma = \{(t,x,y,z)\colon\; y = z = 0\}  \]
in spacetime, and let
\[          D^2 = \{(y,z) \colon\; y^2 + z^2 \le r^2 \} \]
be the disc of radius $r$ in the $yz$ plane.    We describe
some degenerate solutions of general relativity for which $e$ 
vanishes outside the neighborhood $\Sigma \times D^2$ of 
the surface $\Sigma$, and the $BF$ theory equations hold inside this
neighborhood.  

For simplicity, we first consider the case of vanishing cosmological
constant.  Assume that $A$ is of the form
\[      A = A_t dt + A_x dx  \]
inside $\Sigma \times D^2$, with $A_t,A_x$ depending only on $t$ and
$x$.  Let $A$ be arbitrary outside $\Sigma \times D^2$.  
Assume also that $e$ is of form
\[      e = f(y,z)\, [e_y dy + e_z dz]  \]
where $f$ is a smooth function vanishing outside $D^2$
and $e_y, e_z$ depend only on $t$ and $x$.  

Setting 
\[      E = e \we e  ,\]
note that 
\[       E = \eps f^2  \,dy \we dz \]
where 
\[       \eps = e_y \we e_z  \]
is a $\g$-valued function on the surface $\Sigma$.  (Recall that $\g$
is the Lie algebra of $G = \SO(4), \SO(3,1),$ or $\SO(2,2)$, depending on
the signature we are considering.)   Let $\alpha$ denote the restriction of
the connection $A$ to $\Sigma$.  
A calculation then shows that $e$ and $A$ satisfy the equations of
general relativity with vanishing cosmological constant:
\[     d_A(e \we e) = 0 ,\qquad  e \we F = 0, \]
if the following equations hold:
\be          d_\alpha \eps = 0, \qquad \phi = 0,         \label{2dbf} \ee
where $d_\alpha \eps$ is the exterior covariant derivative
of $\eps$ with respect to the connection $\alpha$,
and $\phi$ is curvature of $\alpha$. 
These are precisely the equations
for 2-dimensional $BF$ theory on the surface $\Sigma$.

However, not every solution of 2-dimensional $BF$ theory on $\Sigma$ gives
rise to a degenerate solution of general relativity this way.
Starting from any $\alpha,\eps$ satisfying (\ref{2dbf}), 
we may define the fields $A$ and $E$ as above.  These fields satisfy 
\[         d_A E = 0, \qquad F = 0 \]
inside $\Sigma \times D^2$, and $E$ vanishes outside $\Sigma \times D^2$.
To obtain a solution of general relativity, however, we need a way to write
$E$ as $e \we e$.  There may be no way to do this, or many
ways, but we can certainly do it if $\eps$ is `decomposable', that is, 
if $\eps = e_y \we e_z$ for some sections $e_y,e_z$ of
$\T$.  Note that since $\eps$ is covariantly constant, it will
be decomposable if it is of the form $e_y \we e_z$ at any 
single point of $\Sigma$.  We discuss the issue
of decomposability further in Section \ref{decomposability}.

If the cosmological constant is nonzero things are a bit
more complicated.  Suppose again that $\alpha, \eps$ satisfy 
the 2-dimensional $BF$ theory equations (\ref{2dbf}).  We
set $E = \eps f^2  \,dy \we dz$ as before, but assume
\[      A =  \alpha_t dt + \alpha_x dx + A_\perp \]
inside $\Sigma \times D^2$, with 
\[     A_\perp = A_y dy + A_z dz .\]
Let us show that with an appropriate choice of $A_\perp$,
the fields $A$ and $E$ satisfy the equations of 4-dimensional
$BF$ theory, (\ref{BF}), inside $\Sigma \times D^2$.  It follows 
that if $\eps$ is decomposable, 
we can write $E = e \we e$ and obtain a degenerate solution of the
equations of general relativity with cosmological
constant term, (\ref{einstein}).   

To solve equations (\ref{BF}), it suffices to choose $A_\perp$
such that 
\[     F + {\Lambda \over 6} e \we e = 0 \]
in $\Sigma \times D^2$, or in other words,
\ba  F_{tx} &=& 0      \label{1}  \\
     F_{ty},F_{xy},F_{tz},F_{xz} &=& 0  \label{2} \\  
     F_{yz} &=&-{\Lambda\over 6} \eps.  \label{3} 
\ea
Equation (\ref{1}) holds automatically because $\alpha$ is flat.  
Equations (\ref{2}) say precisely that the $\g$-valued
functions $A_y$ and
$A_z$ are covariantly constant in the $t$ and $x$ directions.
Since $\alpha$ is flat, these equations
have a unique solution given any choice of $A_y$ and $A_z$ on
the disc $\{t = x = 0\} \times D^2$.  We can find
$A_y, A_z$ solving (\ref{3}) on this disc; the solution is not
uniquely, but it is unique up to gauge transformations.  
Choosing any solution and solving equations (\ref{2})
for $A_y$ and $A_z$ on the rest of $\Sigma \times D^2$,
equation (\ref{3}) then holds throughout $\Sigma \times D^2$,
because $F_{yz}$ and $\eps$ are both covariantly constant in 
the $t$ and $x$ directions.

In the next section we discuss the global aspects of these solutions
of general relativity coming from 2-dimensional $BF$ theory,  but
first let us examine their physical significance.   If we restrict one
of our solutions to the slice $t = 0$ we obtain the picture shown in
Figure 1.  Here we have drawn the  `gravitational electric and magnetic
fields' $E$ and $B$, by which we mean  the restriction of the $E$ and
$F$ fields, respectively,  to the slice $t = 0$.   
The field  $E$ vanishes except
in a tube of radius $r$ running along the $x$ axis, so we may think of
our solution as describing a `flux tube' of the gravitational electric
field.  Since $E$ is a $\g$-valued 2-form on space, in Figure
1 we have drawn it as small `surface element' transverse to
the $x$ axis.  Outside the flux tube the gravitational magnetic field is
arbitrary, but inside it we have
\[      B = -{\Lambda \over 6} E \]

\vbox{
\bigskip
\centerline{\epsfysize=1.5in\epsfbox{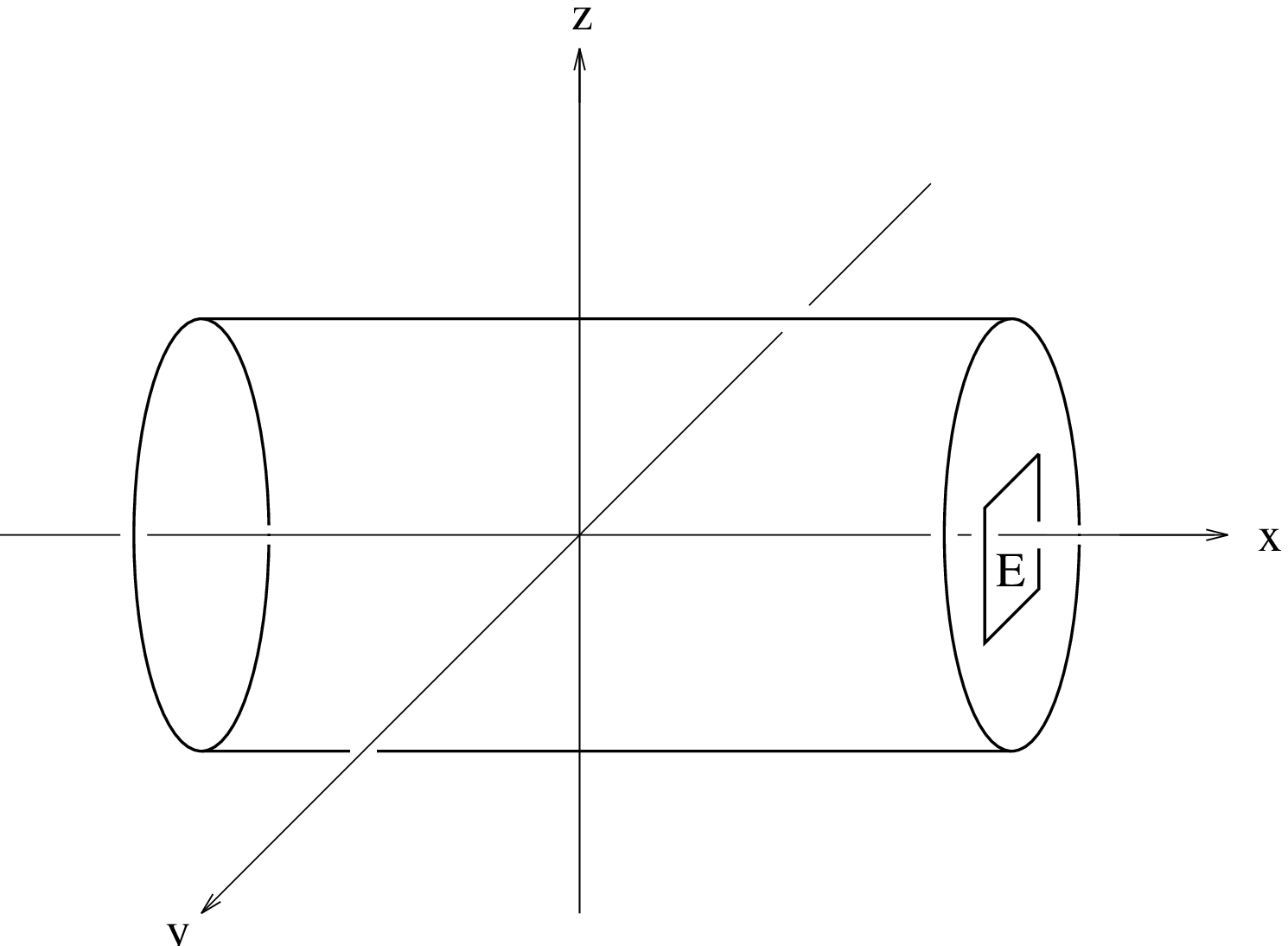}}
\medskip
\centerline{1.  Flux tube of gravitational electric field}
\medskip
}

The 3-metric is zero outside this flux tube, while inside, the
3-metric is degenerate in the $x$ direction but nondegenerate in the
two transverse directions.  In the 3-geometry defined by this
solution, the area of any surface is zero unless it intersects the
flux tube.  We see therefore that this solution is a
kind of classical analog of the `loop states' of quantum gravity,
which have been shown to represent quantized flux tubes of area
\cite{AL,RS2}.  We make this analogy more precise in the next
section, where we consider flux tube solutions associated to arbitrary
thickened links in space.  

The case of nonzero cosmological constant is particularly interesting.
Then $B_{yz}$ is typically nonzero in the flux tube, so the connection
$A$ will typically have nontrivial holonomy around a small loop
wrapping around the tube.  Figure 2 shows a circular loop $\gamma_x$
of radius $r$ centered at the point $x$ on the $x$ axis.  To define
the holonomy around such a loop  as an element of the gauge group $G$, 
rather than a mere conjugacy class, we need to choose a basepoint  for
the loop.  We can do this using a `framing' of the flux tube, or more
precisely, a curve running along the surface of the tube that
intersects each loop $\gamma_x$ at one point, which we take as the
basepoint.  In Figure 2 we have chosen as our framing the line $z =
0, y = -r$.  Using this framing we may define the holonomy
\[      g(x) = T \exp \oint_{\gamma_x} A . \]
Using the definition of parallel transport we easily see that $g(x)$
is covariantly constant along the flux tube:
\[      \partial_x g + [A_x,g] = 0.  \]
In the spacetime picture, $g$ becomes a function of $t$
and $x$, and one can also check that
\[      \partial_t g + [A_t,g] = 0.  \]

\vbox{
\bigskip
\centerline{\epsfysize=1.5in\epsfbox{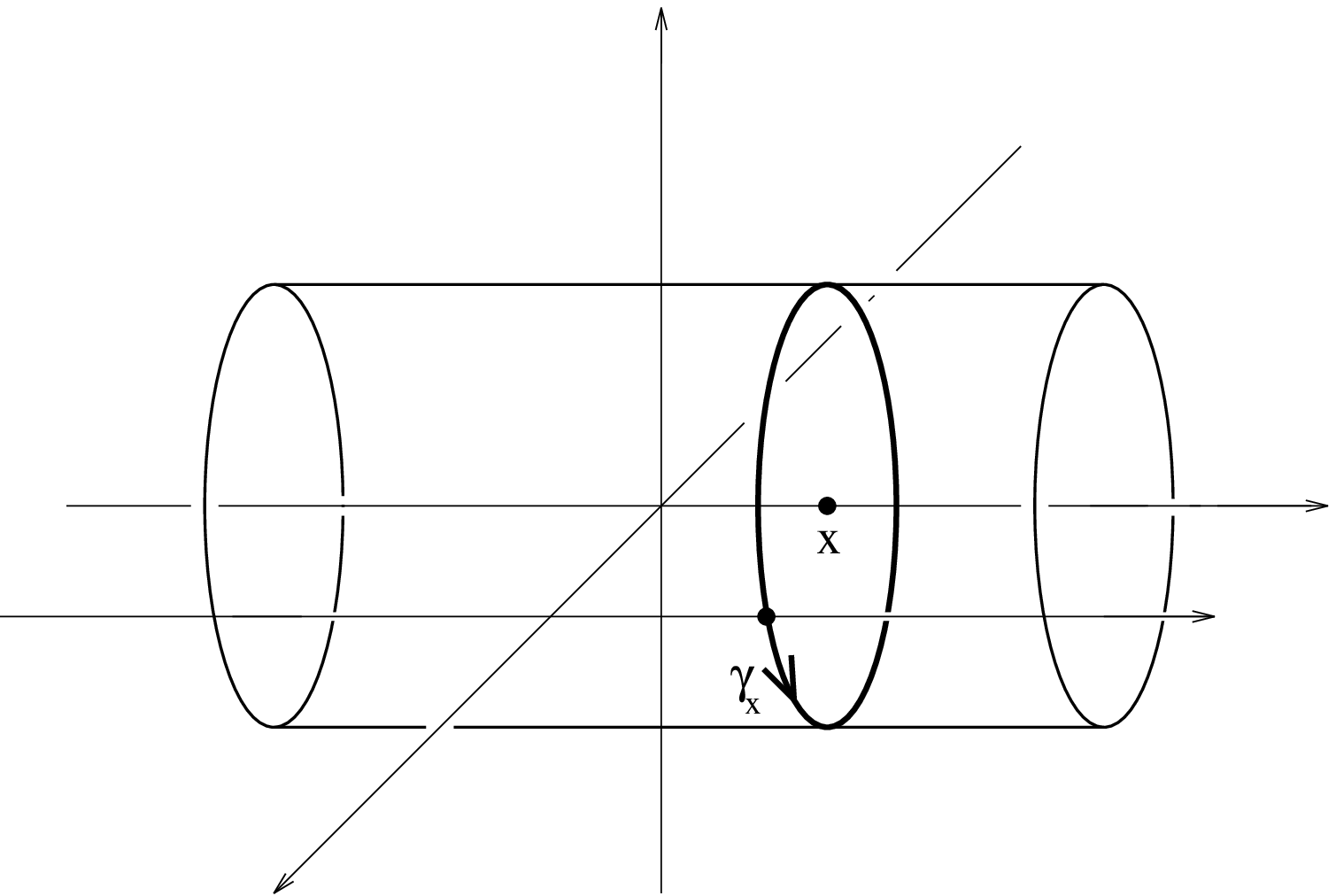}}
\medskip
\centerline{2.  Holonomy of connection around flux tube}
\medskip
}

In the next section we show that in general, we need to equip a
surface $\Sigma$ with the 4-dimensional analogue of a framing
in order to obtain solutions of general relativity from solutions
of 2-dimensional $BF$ theory on $\Sigma$.  This extra structure
also lets us define a field $g$ 
on $\Sigma$ that measures the holonomy of $A$ around small loops
around $\Sigma$.  As above, this field $g$ will be covariantly
constant on $\Sigma$.  In fact, this field is closely related to the
group-valued field in the $G/G$ gauged WZW model on $\Sigma$ 
that we describe in the Conclusions.

\section{Global analysis}

In this section we extend the local analysis of the previous section to
show that for any framed surface embedded in spacetime, we can obtain
degenerate solutions of general relativity from certain solutions of
2-dimensional $BF$ theory on this surface.   
Let $M$ be an oriented 4-manifold, and let $\Sigma$ be a 
2-manifold embedded in $M$.   As before let $\T$ be a vector
bundle isomorphic to $TM$ equipped with a nondegenerate
metric of some fixed signature, and let $P \to M$ be the special
orthogonal frame bundle of $\T$.

By a `framing' of $\Sigma$ we mean a homotopy
class of trivializations of its normal bundle.  
We call an extension of the embedding
\[     i \maps  \Sigma \hookrightarrow M \]
to an embedding 
\[     j \maps  \Sigma \times D^2 \hookrightarrow M \]
a `thickening' of $\Sigma$.  Any thickening determines a framing  in a
natural way.  In fact, any framing comes from some thickening in this
way.  Moreover, two thickenings $j,j'$ determine the same framing if
and only if they are ambient isotopic, that is, if $j' = fj$ for some
diffeomorphism of $M$ connected to the identity.  These results can be
proved just as the analogous results were proved for thickenings of
framed links \cite{B5}.

In what follows we assume $\Sigma$ is equipped with a framing, and we pick
an arbitrary thickening compatible with this framing.  Using this
thickening we think of $\Sigma \times D^2$ as a submanifold of $M$.  We
assume $\alpha$ is a connection on $P|_\Sigma$ and $\eps$ is a section of
$\Ad P$ over $\Sigma$, and we assume that these fields satisfy the equations of
2-dimensional $BF$ theory,
\[          d_\alpha \eps = 0 , \qquad \phi = 0\]
where $\phi$ is the curvature of $\alpha$.   From this data we shall
construct a connection $A$ on $P$ and an  $\Ad P$-valued 2-form $E$ on
$M$ such that $E$ is zero outside the thickening $\Sigma \times D^2
\subseteq M$, and the equations of 4-dimensional $BF$ theory,
(\ref{BF}), hold inside $\Sigma \times D^2$.    

First we construct the field $E$ as follows.  Using notation similar
that of the previous section, let $dy \we dz$ denote the standard
volume form on $D^2$ and let $f$ be a smooth function on $D^2$
vanishing near the boundary.  Pulling back the 2-form $f^2 dy
\we dz$ from $D^2$ to $\Sigma \times D^2$ along the projection
$p_2 \maps \Sigma \times D^2 \to D^2$, we obtain a 2-form on $\Sigma
\times D^2$ which by abuse of language we again denote by $f^2 dy \we
dz$.

We can choose a metric-preserving isomorphism between the bundle
$\T|_{\Sigma \times D^2}$ and the bundle $p_1^\ast T|_\Sigma$, where
$p_1 \maps \Sigma \times D^2 \to \Sigma$ is the projection onto the
first factor.   Requiring that this isomorphism be the identity for
points in $\Sigma$, it is then unique up to gauge transformation.  We
fix such an isomorphism.  Using this, the  pullback $p_1^\ast \eps$
can be thought of as a section of $\Ad P$ on $\Sigma \times D^2$.   We
then define the $\Ad P$-valued 2-form $E$ by 
\[      E = \eps f^2 \, dy \we dz \]
on $\Sigma \times D^2$, and define $E$ to be zero on the rest of $M$.

Next we construct the connection $A$ on $P$.   We pull back $\alpha$
to a connection on $p_1^\ast P|_\Sigma$, and using the isomorphism
chosen in the previous paragraph we think of this as a connection on
$P|_{\Sigma \times D^2}$, which we extend arbitrarily to a connection
$A_{\parallel}$ on all of $P$.   We define $A$ by
\[   A = A_\parallel + A_\perp \]
where $A_\perp$ is an $\Ad P$-valued 1-form given as follows.  
Inside $\Sigma \times D^2$ we set
\[    A_\perp = A_y dy + A_z dz  \]
where $A_z = 0$ and for any point $(p,y,z) \in \Sigma \times D^2$,
\[     A_y(p,y,z) = {\Lambda \over 6} \int_0^z \eps(p,y,z')
f^2(y,z') \; dz' .\]
Outside $\Sigma \times D^2$, we let $A_\perp$ be arbitrary.   

Finally, we check that the equations of 4-dimensional $BF$ theory hold
in $\Sigma \times D^2$.  When $\Lambda \ne 0$ we only need to check
that $F = -{\Lambda\over 6}E$.  We work using local coordinates $t,x$
on $\Sigma$.  Our choice of $A_\perp$ ensures that
\[    F_{yz} = -{\Lambda \over 6} \eps f^2 ,\]
so $F_{yz} = -\Lambda E_{yz}/6$.  We have $E_{tx} = 0$ by construction
and $F_{tx} = 0$ since $A_{\parallel}$ is flat.  We have $E_{ty} = 0$
by construction, and at any point $(p,y,z) \in \Sigma \times D^2$ 
we have 
\ban F_{ty} &=& \partial_t A_y + [A_t,A_y] \\
&=&    {\Lambda \over 6} \int_0^z \Big(\partial_t \eps(p,y,z') +
[ \alpha_t(p),\eps(p,y,z')]\Big) f^2(y,z') \; dz' \\
&=& 0 \ean
since $d_\alpha \eps = 0$.  Similarly we have $E_{tz} = E_{xy} = E_{xz} = 0$
and $F_{tz} = F_{xy} = F_{xz} = 0$.  
When $\Lambda = 0$ we also need to check that $d_A E = 0$ in 
$\Sigma \times D^2$.  This follows from $d_\alpha \eps = 0$.  

Summarizing, we have:

\begin{thm} \et  Let $\Sigma$ be a framed 2-manifold embedded in a
4-manifold $M$, and let $\T$ be a vector bundle over $M$ isomorphic to
the tangent bundle and equipped with a metric $\eta$ of arbitrary
fixed signature.   Let $P$ be the special orthogonal frame bundle of
$\T$.   For any value of $\Lambda$, there is a natural map from
solutions  $(\alpha,\eps)$ of 2-dimensional $BF$ theory on $\Sigma$
(where $\alpha$ is a  connection on $P|_\Sigma$ and $\eps$ is a
section of $\Ad P|_\Sigma$) to gauge and diffeomorphism equivalence
classes of fields $(A,E)$ (where $A$ is a connection on $P$ and $e$ is
a $\Ad P$-valued 2-form on $M$)  satisfying the equations of
4-dimensional $BF$ theory in a neighborhood of $\Sigma$.  Given any of
these solutions, we may extend $A$ and $E$ smoothly to all of $M$ with
$E$ vanishing outside the given neighborhood of $\Sigma$. \end{thm}

Here the term `natural' refers to the fact that given a
diffeomorphism $F \maps M \to M'$ mapping the framed surface $\Sigma
\subseteq M$ to the framed surface $\Sigma' \subseteq M'$ and carrying
the solution $(\alpha,\eps)$ of 2-dimensional $BF$ theory on $\Sigma$ to the
solution $(\alpha', \eps')$ on $\Sigma'$, the corresponding solution
$(A,E)$ of 4-dimensional $BF$ theory on a neighborhood of $\Sigma$ is
mapped to the corresponding solution $(A',E')$ on a neighborhood of
$\Sigma'$, up to diffeomorphism and gauge transformation (or more
precisely, up to a bundle automorphism).

\section{Decomposability} \label{decomposability}

Given the hypotheses of Theorem 1, if $\eps$ is decomposable at one
point of each component of the surface $\Sigma$, it is automatically
decomposable throughout $\Sigma$.  We may thus write $E = e \we e$ and obtain 
a degenerate solution of the equations of general
relativity, (\ref{einstein}), on $M$.  

One way to guarantee decomposability is to 
require that $(\alpha,\eps)$ is
actually a solution of $\SO(3)$ $BF$ theory on $\Sigma$.  Here
we restrict attention to the signatures $++++$ and $+++-$, so that
$G$ is either $\SO(4)$ or $\SO(3,1)$, and we use the fact that
$\SO(3)$ is a subgroup of these groups.  
Suppose we split $\T|_\Sigma$ into a line bundle and a 3-dimensional 
vector bundle on which the internal metric $\eta$ is positive
definite.   Using
a metric-preserving isomorphism between $\T|_{\Sigma \times D^2}$
and $p_1^\ast \T|_\Sigma$, we obtain a splitting
\[       \T_{\Sigma \times D^2} = L \oplus {}^3\T  \]
where $\eta$ is positive definite on ${}^3\T$.  Let ${}^3P$ denote
the special orthogonal frame bundle of ${}^3\T$.  This is a principal
$\SO(3)$-bundle, which we may think of as a sub-bundle of the 
principal $G$-bundle
$P|_{\Sigma \times D^2}$.  Similarly, the vector
bundle $\Ad\, {}^3P$ is a sub-bundle of the vector bundle 
$\Ad P|_{\Sigma \times D^2}$.  Locally, we may think of sections of 
the former bundle as $\so(3)$-valued functions, and sections of 
the latter as $\g$-valued functions, where $\g$
is either $\so(4)$ or $\so(3,1)$.  

Suppose we have a solution of 2-dimensional $BF$ theory on $\Sigma$
for which the connection $\alpha$ reduces to a connection on ${}^3P$ and
the section $\eps$ of $\Ad P$ is actually a section of $\Ad\, {}^3 P$. 
Then $\eps$ is automatically decomposable, since every vector $u \in
\R^3$ is the cross product of two other vectors. 
Moreover, there is a canonical way to decompose $\eps$, since we 
can always write $u = v \times w$ where $v$ and $w$ have the same
length and $u,v,w$ form a right-handed triple.  The map $u \mapsto
(v,w)$ is not smooth at $u = 0$, but if $\eps$ is zero at some point
of $\Sigma$ it is zero throughout that component of $\Sigma$, so
we indeed have a canonical way to write 
\[          \eps = e_y \we e_z  \]
for smooth sections $e_y,e_z$ of ${}^3\T$.

We thus obtain the following:

\begin{thm} \et  Given the hypotheses of Theorem 1, assume that the
signature of $\eta$ is either $++++$ or $+++-$, and assume $\T|_\Sigma$
is equipped with a 3-dimensional sub-bundle on which $\eta$ is positive
definite.  For any value of $\Lambda$, there is a natural map from
solutions  $(\alpha,\eps)$ of 2-dimensional $\SO(3)$ $BF$ theory on $\Sigma$
(where $\alpha$ is a connection on ${}^3P|_\Sigma$ and $\eps$ is a
section of $\Ad\, {}^3P|_\Sigma$) to gauge and diffeomorphism equivalence
classes of fields $(A,e)$ (where $A$ is a connection on $P$ and $e$ is
an $\T$-valued 1-form on $M$) satisfying the equations of general
relativity on $M$.  \end{thm}

One may wonder to what extent these solutions of general relativity
depend on our choice of 3-dimensional sub-bundle of $\T|_\Sigma$,
and whether such a choice always exists.  At least in the case
of signature $++++$, these issues work out nicely.  In this case, we
can always choose a splitting of $\T_\Sigma$ 
into a line bundle $L$ and a 3-dimensional vector bundle $M$,
and this splitting is unique up to a small gauge transformation.  It
follows that our map from solutions $(\alpha,\eps)$ to gauge and
diffeomorphism equivalence classes of solutions $(A,e)$ is independent
of the choice of 3-dimensional sub-bundle of $\T|_\Sigma$, as long as
we choose it so that its orthogonal complement is trivial.

To see these facts we need a little homotopy theory.
Note that a splitting $\T_\Sigma = L \oplus M$ is
equivalent to a reduction of structure group from $\SO(4)$ to
$\SO(3)$.  This is also equivalent to a lifting 
\[
\begin{diagram}
\node[2]{BSO(3)} \arrow{s,r}{j}  \\
\node{\Sigma} \arrow{e,t}{q} \arrow{ne,l}{\ell} \node{BSO(4)} 
\end{diagram}
\]
where $q$ is the classifying map of the bundle $P|_\Sigma$, and $j$ is
the map between classifying spaces induced by the inclusion $\SO(3)
\hookrightarrow \SO(4)$.  Homotopy classes of such liftings correspond
to homotopy classes of maps from $\Sigma$ to $SO(4)/SO(3) \iso S^3$,
which is the homotopy fiber of the map $j$.  However, all maps from
$\Sigma$ to $S^3$ are homotopic, so any two liftings are homotopic.
Thus, suppose we are given two splittings, $\T_\Sigma = L \oplus M$
and $\T_\Sigma = L' \oplus M'$.  These correspond to two liftings
$\ell$ and $\ell'$, and since there exists a homotopy between $\ell$
and $\ell'$, there exists a continuous one-parameter family of
splittings interpolating between the two given ones.  It follows that
there is a small gauge transformation carrying $L$ to $L'$ and $M$ to
$M'$.

\section{Conclusions} \label{conclusions}

Despite many parallels, the relevance --- if any --- of our 
`flux tube' solutions to the loop representation of quantum gravity 
is not yet clear.  Perhaps, however, we can begin to understand this by
quantizing the theory of solutions of this sort associated to a fixed
surface in spacetime.  We can think of this as an unusual sort of
`minisuperspace model'.  In such a model all but finitely many degrees
of freedom of the classical gravitational field are ignored, and then
the rest are quantized.  Presumably any such model is only a
caricature of quantum gravity.  At best we can hope that, like a good
caricature, it reveals certain interesting features of the subject.

Here we give a rough outline of how this might proceed, leaving
a more careful treatment to another paper \cite{B6}.  We follow
a path-integral approach, considering only the
case of nonzero cosmological constant.   Let $M$
be an oriented 4-manifold and let $\T$ be a bundle over $M$
isomorphic to the tangent bundle and equipped with a Lorentzian
metric $\eta$.  Fix a compact framed surface $\Sigma$ in $M$, 
a thickening of $\Sigma$, and a 3-dimensional 
sub-bundle ${}^3\T$ of $\T|_{\Sigma \times D^2}$ as in the previous
section.  Working in the Palatini formalism,
we integrate only over field configurations $(A,e)$ for which:
\begin{enumerate}
\item Inside $\Sigma \times D^2$, $A$ reduces to a connection on 
${}^3\T$ and $F + {\Lambda \over 6} e \we e = 0$.  
\item Outside $\Sigma \times D^2$, $A$ is arbitrary and $e = 0$.
\end{enumerate}
For such field configurations the action can be expressed purely
in terms of $A$:
\ban  S_{Pal}(A,e) &=& \int_M \tr(e \we e \we F + {\Lambda \over 12}
e \we e \we e \we e) \\
&=& - {3\over \Lambda} \int_{\Sigma \times D^2} \tr(F \we F) \ean
If we fix a connection $A_0$ of the above form that is flat on
the boundary $\Sigma \times S^1$ of the thickening, and use this 
to think of $A$ as an $\Ad P$-valued 1-form, we obtain by Stokes' theorem
\[   S_{Pal}(A,e) = - {3\over \Lambda} S_{CS}(A) + c  \]
where $c$ depends only on the fixed connection $A_0$ and $S_{CS}(A)$ is
the Chern-Simons action 
\[   S_{CS}(A) = \int_{\Sigma \times S^1} 
\tr(A \we dA + {2\over 3} A \we A \we A)  \]


Now we are in a position to take full
advantage of the following `ladder of field theories' \cite{B2} 
in dimensions 2, 3, and 4: 

\vbox{ 
\vskip 2em
\small
\noindent \hskip 1em
{\bf 4d:}  \hskip 3em {\bf General relativity}
\hskip 1em $\rightarrow$ \hskip 1em {\bf $BF$ theory}
\hskip 1em $\rightarrow$  \hskip 1em {\bf Chern theory}
\vskip 1em
\noindent \hskip 2em
\hskip 28em $\downarrow$
\vskip 1em
\noindent \hskip 1em
{\bf 3d:} \hskip 21em {\bf Chern-Simons theory }
\vskip 1em
\noindent  \hskip 2em
\hskip 28em $\downarrow$  
\vskip 1em
\noindent \hskip 1em 
{\bf 2d:} \hskip 8em {\bf WZW model} \hskip 2em 
$\rightarrow$ \hskip 2em {\bf $G/G$ gauged WZW model} 
\vskip 2em  }

\noindent As we have seen, for any framed surface $\Sigma$ in spacetime, 
general relativity has a class of degenerate solutions coming from
4-dimensional $BF$ theory.  Our minisuperspace model amounts to
4-dimensional $BF$ theory on the thickened surface $\Sigma \times
D^2$.  Using the equations of motion of $BF$ theory to write the
action purely as a function of $F$, it is then proportional to the
second Chern number
\[   c_2 = {1\over 8 \pi^2} \int_M \tr(F \we F).  \]
This action gives a theory involving only the $A$ field, which
we may call `Chern theory'.   The Lagrangian
of Chern theory is a closed 4-form, so it has no bulk 
degrees of freedom, and on $\Sigma \times D^2$ it reduces to
Chern-Simons theory on $\Sigma \times S^1$.   

The final step down the dimensional ladder uses the fact that
Chern-Simons theory on a 3-manifold of form $\Sigma \times S^1$ is
equivalent to the $G/G$-gauged Wess-Zumino-Witten model on $\Sigma$.
This theory is a 2-dimensional topological field theory
having a basis of states corresponding to irreducible representations
of the quantum group associated to the gauge group $G$, which in our
case is $\SO(3)$.  We see therefore that in the canonical picture,
when $\Sigma$ intersects space a given time in some link $L$, states
of our minisuperspace model are described to labellings of the
components of $L$ with irreducible representations of quantum
$\SO(3)$, or in other words, integral spins $j = 0, 1, 2, \dots, k/2$
where the `level' $k$ is an even integer depending on $\Lambda$.  A
link labelled this way is simply a $q$-deformed spin network with no
vertices.

While we have included the Wess-Zumino-Witten model in our picture
of the ladder of field theories, it has only an indirect relevance
to our minisuperspace model.  It is a conformal field theory whose
conformal blocks form a basis of states of the $G/G$ gauged
Wess-Zumino-Witten model.  In string theory, the string
worldsheet is equipped with a conformal structure, and some of the
transverse vibrational modes of the string are described by
Wess-Zumino-Witten models living on this surface.  In the
minisuperspace model sketched above, we see instead the $G/G$ gauged
Wess-Zumino-Witten model living on the surface $\Sigma$.  There is, of
course, no opportunity for us to see a conformal field theory on
$\Sigma$, because this surface is not equipped with a metric; even in
any particular solution the metric is only nondegenerate in the two
directions transverse to $\Sigma$.  

This begins to make more precise
our previous speculation that the loop representation of quantum
gravity is related to background-free string theory, with the loops
arising as $t = 0$ slices of string worldsheets \cite{B}.  It is
intriguing that in a somewhat different framework, Jacobson
\cite{J} has constructed degenerate solutions of general
relativity in which the metric is nondegenerate in the directions
tangent to a surface, and by this means obtains a conformal field
theory.  Comparing his solutions and ours may shed more light on 
the relation between the loop representation of quantum gravity
and string theory --- or at least 2-dimensional field theory.  

\subsection*{Acknowledgements} 

I would like to thank Ted Jacobson for getting me interested in
degenerate solutions of general relativity coming from field theories
on surfaces, Mike Reisenberger for pointing out that his `2-sphere
solution' of general relativity should be related to $BF$ theory, 
Jorge Pullin for emphasizing the importance of decomposability of the
$E$ field, and W.\ Dale Hall and Clarence Wilkerson for helping me with
the homotopy theory.  I would also like to thank the Erwin Schr\"odinger
Institute in Vienna, the Theoretical Physics Group at Imperial
College, and the Center for Gravitational Physics and Geometry at
Pennsylvania State University for their hospitality while this work
was being done.

\end{document}